\documentclass{camera}

\def\ep{\epsilon}

\newcommand\as{\alpha_{\mathrm{S}}}

\begin{document}
\renewcommand{\thefootnote}{\fnsymbol{footnote}}
%
\title{PROGRESS IN PERTURBATIVE QCD CALCULATIONS FROM LEP TO TEVATRON RUN II AND LHC
$^*$
}
%
\author{M. Grazzini}

\vspace*{-1cm}

\organization{Dipartimento di Fisica, Universit\`a di Firenze\\ and\\
INFN, Sezione di Firenze, Largo Fermi 2, I-50125 Firenze, Italy}

\maketitle

\vspace*{-1cm}

\abstract{I briefly review the progress in perturbative QCD calculations in LEP era and discuss the prospects for the extension of their accuracy to higher orders.}
%
\footnotetext[1]{Invited talk given at the XIII italian meeting on high energy physics ``LEPTRE'', Rome April 18-20 2001, to appear in the proceedings.}

\renewcommand{\thefootnote}{\arabic{footnote}}

\vskip 1cm

Higher order QCD calculations are essential in order to assess and precisely quantify our confidence in the Standard Model. Perturbative predictions at the Leading Order (LO) in the QCD coupling $\as$ rely
on tree-level matrix elements and
therefore
provide only an approximate description of cross sections and distributions. The unphysical
dependence on
renormalization and factorization scales
turns out to be quite large.

The simplest higher order calculations are those for
fully
inclusive observables. Examples of these quantities are $\sigma({\rm e^+e^-\to hadrons})$, the hadronic branching ratio of the $Z$ and of the $\tau$ lepton. For these quantities infrared (IR) singularities essentially cancel at the integrand level and thus accurate predictions exist up to next-to-next-to-leading order (NNLO) \cite{inclusive}.
Moreover these observables
are affected by
small non perturbative corrections and
thus they are particularly suitable for $\as$ measurements \cite{Dissertori}.
 
The most detailed QCD studies at $e^+e^-$ colliders are based on {\em event shapes} and {\em jet cross sections}.
Event shapes variables are quantities that characterize the structure of the hadronic event. 
Thus, with respect to fully inclusive observables, event shape distributions are useful not only to measure $\as$ but also to perform stringent QCD tests. The price to pay is that non perturbative effects are more important \cite{Banfi} and that the perturbative calculation is more difficult.
Real and virtual contributions have a different number of final state partons and they should be integrated separately in order to cancel IR singularities. This is only possible by combining analytical techniques with numerical (Monte Carlo) integration. In practice one performs analytically only the part of the integration that produces the singularities. There are two methods of this kind: the subtraction method \cite{Ellis:1981wv} and the slicing method \cite{Fabricius:1981sx}\footnote{A complete numerical method was proposed in Ref.\cite{Soper:1998ye}.}.
The NLO calculation for all the relevant $3$-jet observables was performed by using the subtraction method in Ref.\cite{Kunszt:1989km}.

At the beginning of LEP era the standard procedure was the comparison of the data with NLO QCD.
Far away from the two-jet region the NLO calculation with
the renormalization scale $\mu$ of order of the
center-of-mass
energy $Q$ usually
gives good fits to the data. On the contrary,
very small (and unphysical) values of $\mu$ \cite{Dissertori}
are required in order
to extend the range of the fit.

This is a signal of the fact that close to the boundary of the phase space (two-jet region) large logarithmic corrections of the form $\as \log^2 1/y$ appear that spoil the perturbative expansion
(here $y$ denotes the variable that becomes small in the two-jet region).
These contributions, whose origin is due to the emission of soft and collinear gluons,
have to be resummed to all orders.
The resummation formalism was developed up to next-to-leading-logarithmic (NLL) accuracy for those shape variables that fulfill {\em exponentiation} \cite{shape}. This property
is a consequence of two basic conditions:
$i)$ {\em matrix element factorization}; $ii)$ {\em phase space factorization}.
The first is a {\em dynamical} condition that relies on the general properties of soft and collinear emission. The second is a {\em kinematical} condition that depends on the particular quantity we consider. Both have to be satisfied in order to be able to perform resummation. The NLL resummed calculation is matched with the NLO prediction (NLL+${\cal O}(\as^2)$) to give a result that is everywhere as good as the NLO result and much better in the $y\to 0$ region. 
Extensive studies based on NLL+${\cal O}(\as^2)$ calculations have been carried out over the past years \cite{Dissertori}.
These studies have shown that resummed predictions are less sensitive to
the renormalization scale $\mu$ and in particular
enable us
to avoid the choice
of extremely small (unphysical) $\mu$ in the two-jet region.

A jet is qualitatively defined as a collimated spray of energetic hadrons, and is considered as a typical signal of parton dynamics at short distances. However, in order to perform quantitative studies, one needs a precise definition of jet.
Once a jet definition has been chosen, the $n-$jet cross section is a function that depends only on the resolution parameter $y_{cut}$.
When $y_{cut}$ becomes small jet cross sections develop large $\as\log^2 1/y_{cut}$  corrections
similar to the ones that affect shape distributions in the two-jet region.
As it happens for shape variables, the resummation of these corrections is possible if conditions $i)$ and $ii)$ above are satisfied.
The old JADE algorithm \cite{Bartel:1986ua} does not allow to perform the resummation since it
induces
strong kinematic correlations that spoil $ii)$. The $k_T$ algorithm was introduced to make the resummation in the $y_{cut}\to 0$ region possible \cite{Catani:1991hj}.
NLL+${\cal O}(\as^2)$ predictions exist for jet rates \cite{Catani:1991hj}
and
jet
multiplicities \cite{Catani:1992pm}.

As far as NLO calculations are concerned, there
were important developments
in the last few years. In the first applications both the subtraction
and the slicing methods were used with extensive partial fractioning on the QCD matrix element.
This procedure is intrinsically process dependent.
Later it was understood that both methods can be generalized in a process
independent manner. The key observation is that the IR singularities
can be singled out in a universal manner by using the factorization
properties of soft and collinear emission.
Today we have general algorithms that in principle allow to compute any observable at NLO, both with the subtraction \cite{sub,Catani:1997vz} and with the slicing method \cite{sli}.
As far as LEP physics is concerned new calculations appeared for: $3$-jet observables \cite{Catani:1997vz}\footnote{With respect to the calculation of Ref.\cite{Kunszt:1989km} this one includes the contribution of the $Z$ and allows to study the orientation of the event.}, $3$-jet with mass effects included \cite{threejetm} and $4$-jet \cite{fourjet}.

With this situation one may wonder why we should do an effort to go to NNLO.
There are several reasons to do that. The first one is that since reliable perturbation theory starts at NLO, error estimate should start at NNLO.  As far as LEP is concerned, the NNLO calculation of $3$-jet observables would considerably reduce the error on $\as$ \cite{Dissertori}. Then we should consider that improved experimental techniques and higher luminosities will require a better control on the QCD background. 

A NNLO calculation requires in general three basic ingredients:
$~i)$ NNLO parton distributions;
$~ii)$ two-loop amplitudes;
$~iii)$ knowledge of the IR behaviour of tree-level and one-loop amplitudes at ${\cal O}(\as^2)$.

In the last years enormous progress has been achieved in all these aspects.
For a consistent evaluation of a NNLO cross section at hadron colliders NNLO (three loop) parton distributions are required. Even though their NNLO evolution kernels are
not fully available, some of their Mellin moments have been computed
\cite{vermaseren} and,
from these, approximated kernels have been constructed \cite{vnvogt}.
Recently, the new MRST \cite{mrst2000} set of distributions became
available, including the (approximated) NNLO 
densities, which allows an evaluation of the hadronic
cross section to (almost full) NNLO accuracy.

Until two years ago no two-loop amplitude depending on more than one scale was available.
Thanks to the calculations of important two-loop master diagrams \cite{dbox} and to
the great progress in the reduction of tensor integrals \cite{Anastasiou:2000mf}
the first calculations of two-loop $2\to 2$ amplitudes have recently appeared in QED \cite{Bern:2001ie},
and in QCD \cite{glover}. 
The IR singularities appearing in these amplitudes are in agreement with the general
prediction in  Ref.\cite{Catani:1998bh}.
These results will be relevant for the calculation of Bhabha scattering
and two-jet cross sections in hadron collisions at NNLO.

To perform a NNLO calculation one has to combine the two-loop amplitude with the
one-loop correction where one parton is unresolved and with the tree-level contribution
where two partons are unresolved.
The kernels that control soft \cite{Bern:1998sc,Catani:2000pi}
and collinear \cite{Bern:1998sc,Kosower:1999rx}
singularities appearing in one-loop amplitudes have been computed.
The IR singularities appearing in tree-level amplitudes are more complicated because
many soft/collinear limits have to be considered. All these limits have been studied
\cite{Campbell:1998hg,Catani:2000ss}
and the corresponding kernels have been computed \cite{Campbell:1998hg}--\cite{DelDuca:2000ha}.

The step that remains to be performed is to combine all
these ingredients
to construct general algorithms to handle and cancel infrared singularities.
This step is more difficult than at NLO since the pattern of IR singularities is much more complicated. Nevertheless some applications where some progress in this direction has been achieved recently appeared \cite{Gehrmann-DeRidder:1998gf,deFlorian:2000pr,Catani:2001ic}.

The results of
Refs.\cite{Bern:1998sc}--\cite{DelDuca:2000ha}
are relevant not only to perform NNLO calculations, but also to extend
the accuracy of resummed calculations at NNLL.
Transverse momentum ($k_T$) distributions of high-mass systems (lepton pairs, vector boson, Higgs...) in hadronic collisions
are affected in the small $k_T$ limit by large logarithmic contributions of the same (infrared) nature of the ones present in
event-shape distributions in the two-jet limit.

In Ref.\cite{deFlorian:2000pr} the structure of these large corrections was studied at ${\cal O}(\as^2)$ up
to NNLL accuracy. This calculation was performed in a general (process independent)
manner by exploiting the universal nature of these corrections.
The results of Refs.\cite{Bern:1998sc}--\cite{Catani:1999nv}
were used
to construct improved approximations of the relevant matrix elements that allow to control
all the infrared singular regions responsible for the appearance of the logarithmic contributions.
This method, even if strongly dependent on the special kinematics of this class of processes, could be extended
in the spirit of Ref.\cite{Catani:1997vz} to more general cases
\footnote{The results of Ref.\cite{deFlorian:2000pr}, combined with the numerical estimate \cite{vnvogt}
of the  coefficient that controls the soft-collinear singularity in the three-loop splitting functions
will allow a (partial) extension of the accuracy
of these resummed calculations at NNLL.}. 

In Ref.\cite{Catani:2001ic} the calculation of the soft and virtual
corrections to Higgs boson production at hadron colliders was presented.
This calculation was done by combining the recent
results \cite{Harlander:2000mg} for the two-loop amplitude $gg\to H$
in the large $m_{top}$ limit with the soft factorization formulae for tree-level
\cite{Campbell:1998hg,Catani:2000ss} and one-loop \cite{Bern:1998sc,Catani:2000pi}
amplitudes
\footnote{The same calculation was independently
performed in Ref.\cite{Harlander:2001is}
by means of the direct evaluation of the relevant
amplitudes in the soft limit.}.
From the theoretical side this calculation is very important since it provides
a check of the cancellation of the IR poles from $1/\ep^4$ to $1/\ep$ between real and
virtual contributions.
From the phenomenological side the results give a first consistent estimate of
the QCD corrections to this important process at NNLO.

Up to a few years ago NNLO calculations, if doable,
were considered very far in the future. With the progress
achieved
in the recent years we can be more confident that these calculations will be
feasible in the LHC era.

\noindent {\bf Acknowledgments.} I wish to thank Stefano Catani and G\"unther Dissertori for their help in preparing this talk.


\begin{thebibliography}{90}


\bibitem{inclusive}
S.~G.~Gorishnii, A.~L.~Kataev and S.~A.~Larin,
Phys.\ Lett.\ B {\bf 259} (1991) 144;
L.~R.~Surguladze and M.~A.~Samuel,
Phys.\ Rev.\ Lett.\  {\bf 66} (1991) 560
[Erratum-ibid.\  {\bf 66} (1991) 2416].


\bibitem{Dissertori}
G. Dissertori, these proceedings.

\bibitem{Banfi}
A. Banfi, these proceedings.

\bibitem{Ellis:1981wv}
R.~K.~Ellis, D.~A.~Ross and A.~E.~Terrano,
Nucl.\ Phys.\ B {\bf 178} (1981) 421.

\bibitem{Fabricius:1981sx}
K.~Fabricius, I.~Schmitt, G.~Kramer and G.~Schierholz,
Z.\ Phys.\ C {\bf 11} (1981) 315.


\bibitem{Soper:1998ye}
D.~E.~Soper,
Phys.\ Rev.\ Lett.\ {\bf 81} (1998) 2638.

\bibitem{Kunszt:1989km}
Z.~Kunszt, P.~Nason, G.~Marchesini and B.~R.~Webber,
in ``Z Physics at LEP 1'', CERN 89-08, vol. 1, p. 373.


\bibitem{shape}
S.~Catani, L.~Trentadue, G.~Turnock and B.~R.~Webber,
Nucl.\ Phys.\ B {\bf 407} (1993) 3.

\bibitem{Bartel:1986ua}
W.~Bartel {\it et al.}  [JADE Collaboration],
Z.\ Phys.\ C {\bf 33} (1986) 23.

\bibitem{Catani:1991hj}
S.~Catani, Y.~L.~Dokshitzer, M.~Olsson, G.~Turnock and B.~R.~Webber,
Phys.\ Lett.\ B {\bf 269} (1991) 432.

\bibitem{Catani:1992pm}
S.~Catani, Y.~L.~Dokshitzer, F.~Fiorani and B.~R.~Webber,
Nucl.\ Phys.\ B {\bf 377} (1992) 445.


\bibitem{sub}
S.~Frixione, Z.~Kunszt and A.~Signer,
Nucl.\ Phys.\ B {\bf 467} (1996) 399;
Z.~Nagy and Z.~Trocsanyi,
Nucl.\ Phys.\ B {\bf 486} (1997) 189.


\bibitem{Catani:1997vz}
S.~Catani and M.~H.~Seymour,
Nucl.\ Phys.\ B {\bf 485} (1997) 291
[Erratum-ibid.\ B {\bf 510} (1997) 291].


\bibitem{sli}
W.~T.~Giele and E.~W.~Glover,
Phys.\ Rev.\ D {\bf 46} (1992) 1980;
W.~T.~Giele, E.~W.~Glover and D.~A.~Kosower,
Nucl.\ Phys.\ B {\bf 403} (1993) 633;
S.~Keller and E.~Laenen,
Phys.\ Rev.\ D {\bf 59} (1999) 114004.


\bibitem{threejetm}
W.~Bernreuther, A.~Brandenburg and P.~Uwer,
Phys.\ Rev.\ Lett.\ {\bf 79} (1997) 189;
P.~Nason and C.~Oleari,
Nucl.\ Phys.\ B {\bf 521} (1998) 237;
G.~Rodrigo, M.~Bilenky and A.~Santamaria,
Nucl.\ Phys.\ B {\bf 554} (1999) 257.


\bibitem{fourjet}
L.~Dixon and A.~Signer,
Phys.\ Rev.\ D {\bf 56} (1997) 4031; 
Z.~Nagy and Z.~Trocsanyi,
Phys.\ Rev.\ Lett.\ {\bf 79} (1997) 3604;
S.~Weinzierl and D.~A.~Kosower,
Phys.\ Rev.\ D {\bf 60} (1999) 054028.

\bibitem{vermaseren}
S. A. Larin, P. Nogueira, T. van Ritbergen and J. A. M. Vermaseren,
Nucl. Phys.\ B {\bf 492} (1997) 338;
A. Retey and J. A. M. Vermaseren, hep-ph/0007294.


\bibitem{vnvogt}
W.~L.~van Neerven and A.~Vogt,
Nucl.\ Phys.\ B {\bf 568} (2000) 263,
Nucl.\ Phys.\ B {\bf 588} (2000) 345.

\bibitem{mrst2000}
A.~D.~Martin, R.~G.~Roberts, W.~J.~Stirling and R.~S.~Thorne,
Eur.\ Phys.\ J.\ C {\bf 18} (2000) 117.

\bibitem{dbox}
V.~A.~Smirnov,
Phys.\ Lett.\ B {\bf 460} (1999) 397;
J.~B.~Tausk,
Phys.\ Lett.\ B {\bf 469} (1999) 225.

\bibitem{Anastasiou:2000mf}
C.~Anastasiou, T.~Gehrmann, C.~Oleari, E.~Remiddi and J.~B.~Tausk,
Nucl.\ Phys.\ B {\bf 580} (2000) 577.

\bibitem{Bern:2001ie}
Z.~Bern, L.~Dixon and A.~Ghinculov,
Phys.\ Rev.\ D {\bf 63} (2001) 053007.


\bibitem{glover}
C.~Anastasiou, E.~W.~Glover, C.~Oleari and M.~E.~Tejeda-Yeomans,
hep-ph/0010212,
hep-ph/0011094,
hep-ph/0101304;
E.~W.~Glover, C.~Oleari and M.~E.~Tejeda-Yeomans,
hep-ph/0102201.


\bibitem{Catani:1998bh}
S.~Catani,
Phys.\ Lett.\ B {\bf 427} (1998) 161.



\bibitem{Bern:1998sc}
Z.~Bern, V.~Del Duca and C.~R.~Schmidt,
Phys.\ Lett.\ B {\bf 445} (1998) 168;
Z.~Bern, V.~Del Duca, W.~B.~Kilgore and C.~R.~Schmidt,
Phys.\ Rev.\ D {\bf 60} (1999) 116001.

\bibitem{Catani:2000pi}
S.~Catani and M.~Grazzini,
Nucl.\ Phys.\ B {\bf 591} (2000) 435.


\bibitem{Kosower:1999rx}
D.~A.~Kosower and P.~Uwer,
Nucl.\ Phys.\ B {\bf 563} (1999) 477.


\bibitem{Campbell:1998hg}
J.~M.~Campbell and E.~W.~Glover,
Nucl.\ Phys.\ B {\bf 527} (1998) 264.


\bibitem{Catani:2000ss}
S.~Catani and M.~Grazzini,
Nucl.\ Phys.\ B {\bf 570} (2000) 287.



\bibitem{Catani:1999nv}
S.~Catani and M.~Grazzini,
Phys.\ Lett.\ B {\bf 446} (1999) 143.

\bibitem{DelDuca:2000ha}
V.~Del Duca, A.~Frizzo and F.~Maltoni,
Nucl.\ Phys.\ B {\bf 568} (2000) 211.



\bibitem{Gehrmann-DeRidder:1998gf}
A.~Gehrmann-De Ridder and E.~W.~Glover,
Nucl.\ Phys.\ B {\bf 517} (1998) 269.


\bibitem{deFlorian:2000pr}
D.~de Florian and M.~Grazzini,
Phys.\ Rev.\ Lett.\ {\bf 85} (2000) 4678.

\bibitem{Catani:2001ic}
S.~Catani, D.~de Florian and M.~Grazzini,
JHEP {\bf 0105} (2001) 025.


\bibitem{Harlander:2000mg}
R.~V.~Harlander,
Phys.\ Lett.\ B {\bf 492} (2000) 74.

\bibitem{Harlander:2001is}
R.~V.~Harlander and W.~B.~Kilgore,
hep-ph/0102241.














\end{thebibliography}
\end{document}